\documentclass[a4paper]{article}
\pdfoutput=1
\usepackage[margin=25mm]{geometry}
\usepackage{amsmath}
\usepackage{amsfonts}
\usepackage{amssymb}
\usepackage{graphicx}
\usepackage{verbatim}
\immediate\write18{texcount -tex -sum  \jobname.tex > \jobname.wordcount.tex}
\usepackage{subfigure,epsfig,amsfonts,amsmath,cases,amssymb,graphics, latexsym, times, algorithmic, algorithm, bbm, color, soul,wrapfig,comment}
\usepackage{url}

\providecommand{\keywords}[1]
{
	\small	
	\textbf{\textit{Keywords---}} #1
}

\title{Optimal Scheduling of Anticipated COVID-19 Vaccination: A Case Study of New York State }
\author{Syed Irfan Ali Meerza\thanks{Syed Irfan Ali Meerza is a Ph.D student at the Department of Electrical and Computer Engineering, University of Louisville, Louisville, KY, USA}, 
	Seyed M. Karimi\thanks{Seyed M. Karimi is a Assistant Professor at the Department of Health Management and System Sciences, University of Louisville, Louisville, KY,USA.}, 
	Bert B. Little\thanks{Bert B. Little is a Professor at the Department of Health Management and System Sciences, University of Louisville, Louisville, KY,USA.},\\ 
	Jacek M. Zurada\thanks{Jacek M. Zurada is a Professor at the Department of Electrical and Computer Engineering, University of Louisville, Louisville, KY,USA.}, 
	Tamer Inanc\thanks{Tamer Inanc is a Associate Professor at the Department of Electrical and Computer Engineering, University of Louisville, Louisville, KY,USA.} \\
}
\date{July, 2020} % Comment this line to show today's date

\begin{document}
	\maketitle
	
	\begin{abstract}
		This study aims to  determine an optimal control strategy for vaccine scheduling in COVID-19 pandemic treatment by converting widely acknowledged infectious disease model named SEIR into an optimal control problem. The problem is augmented by adding medication and vaccine limitations to match real-world situations. Two version of the problem is formulated to minimize the number of infected individuals at the same provide the optimal vaccine possible to reduce the susceptible population to a considerably lower state. Optimal control problems are solved using RBF-Galerkin method.\:These problems are tested with a benchmarking dataset to determine required parameters. After this step, problems are tested with recent data for New York State, USA. The results regarding the proposed optimal control problem provides a set of evidences from which an optimal strategy for vaccine scheduling can be chosen, when the vaccine for COVID-19 will be available.
	\end{abstract} \hspace{10pt}
	
	%TC:ignore
	\section{Introduction}
	\label{sec:intro}
	
	From the starting of the year 2020, the world faced a great challenge of the coronavirus outbreak. It started in December 2019 when there was a sudden increase of pneumonia of unknown etiology reported in Wuhan city, Hubei Province, China~\cite{huang2020clinical}~\cite{lu2020out}. A scientific team of the Chinese Academy of Engineering announced that a new coronavirus has caused this outbreak. This new coronavirus was named as SARS-CoV-2 by the International Committee on Taxonomy of Viruses (ICTV) for its similarities with the virus SARS-CoV found in 2003~\cite{ictv}. This new virus causes acute respiratory illness and even fatal acute respiratory distress syndrome (ARDS). The diseases associated with SARS-CoV-2 was named COVID-19 by the World Health Organization (WHO)~\cite{who}. The virus spread rapidly throughout the Hubei province. Researchers had confirmed, the virus can spread through human-to-human transmission~\cite{li2020early}. As a person exposed to the virus can take 12 days to grow symptoms~\cite{lauer2020incu} and also due to globalization and easy transportation options the virus spread quickly throughout the world and created a pandemic situation. As of May 12, 2020, according to worldometer data 4,306,378 people are infected by the coronavirus among them 292,946 people are died by COVID-19 disease~\cite{worldometer}.
	
	The pandemic of coronavirus caused a crisis in public health accompanying social fear and economic fall in different countries. Due to the fast spread of SARS-CoV-2 in various parts of the world WHO declared it as Public Health Emergency of International Concern (PHEIC) and the global level assessment of the risk of the impact of COVID-19 is set to very high~\cite{who2}. Vaccination is the principal long term defense measure from any infectious disease. There is more than $100$ vaccines are under development with some of them in human trial~\cite{CNNVaccine}. Nine pharmaceutical giants are racing for to introduce the vaccine for COVID-19. However, having a vaccine is only the beginning for this kind of large outbreaks. After the introduction of the vaccine main challenge is the distribution of the possibly limited number of future vaccine supplies to the large global population. That's why the optimization of the vaccination policies before their implementation is necessary to allocate resources in a better way. The optimal use of vaccines depends on various factors including the structure of the population, vaccine availability, and transportation availability levels~\cite{kim2016constrained}.
	
	Several researchers have attempted to develop optimal strategies for different infectious diseases including COVID-19. Kim et al. applied constrained optimal control for vaccination of influenza~\cite{kim2016constrained}. In~\cite{Pinho2014Optimal} de Pinho et al. used an optimal control approach to SEIR model and introduced l\textsubscript{1} cost which is linear with respect to the control variable to the model and also developed mixed constraints for the model. They limited the number of vaccines available for the policy and solved the model. De Pinho et al.~\cite{Pinho2016costs} normalized the SEIR model and tried three different strategies  and analyzed their cost. They also discussed the pros and cons of the normalized SEIR model and compared it with the classical model~\cite{Pinho2016application}. Libottea et al. took the optimal control approach toward the COVID-19 and used mono and multi-objective optimal control to determine the optimal policy for the vaccine administration in COVID-19 treatment~\cite{libotte2020determination}. The mono objective minimizes the infected during the time horizon while the multi-objective control minimizes the infected individuals and the prescribed vaccine concentration. They have considered real data from china in their model. 
	
	Above mentioned works have focused on the solution of vaccine distribution or scheduling. Only some of them has considered the vaccine availability during the distribution. Some of them have considered to introduce the medicated individual in a different group. Almost all of them used hypothetical data for solution. In our work we have considered medicated individuals in the model as they have different death and recovery rate from the other infected ones. To make the model more real world worthy we have limited the everyday vaccine availability, as newly introduced vaccine will not be in unlimited supply and it will be a challenge to provide the vaccine to different locations in a short time. We solved the models for the hypothetical data used by the work mentioned above to compare the results and also solved the model for real world data of New York State, USA.
	
	In this paper, we used normalized SEIR model with added medication compartment called SEIRM model and solve for two different initial data, one is a synthesized data and another is real data for New York State, USA. We have used these data to test the SEIRM model with and without vaccine limitation. We have used RBF-Galerkin method, recently developed by Hossein et al. in~\cite{Mirinejad2016RBF} to find optimal vaccine scheduling. RBF-Galerkin uses Radial Basis Function (RBF) approximation of states and controls along with Galerkin error projection and collocate them on random nodes to translate the optimal control problem into a nonlinear programming (NLP) problem foe numerical solution. 
	
	%Summarize the existing work and articulate our novelty.
	
	We summarize our contributions as follows:
	\begin{enumerate}
		\item We have used a SEIR model with added compartment called the medicated which contains the individuals who are being medicated for the COVID-19 to better suit the current situation of the pandemic. We have also added limitation on the supply for vaccine for each day mimic one of the most important hurdle in the distribution. 
		
		\item We have used the RBF-Galerkin method to solve the model. RBF-Galerkin method has locality property, which makes it numerically stable as noise will not influence all the coefficients in the approximation process. We have solved the model with and without vaccine limitation to compare the results.
		
		\item These models are solved two different datasets, one is the synthesized data used by many researchers to benchmark our solution then we used a dataset containing real world data of New York state, USA taken on May 12, 2020. 
		
	\end{enumerate}
	The remainder of this paper is organized as follows. First, we discuss the mathematical models in Section~\ref{sec:math}. Then, the optimal control problem related to the mathematical models is presented in Section~\ref{sec:optimal}. In Section~\ref{sec:RBF} we have introduced the RBF-Galerkin method for solving optimal control problems and also the solution of the optimal control problem using the RBF-Galerkin method. In Section~\ref{sec:numerical solution}, we have solved the optimal vaccination distribution model using real-world data of New York State,USA. Finally, our conclusions are given in Section~\ref{sec:conclusion}.
	
	\section{Mathematical Model}
	\label{sec:math}
	Researchers have used various compartmental models to represent the evaluation of epidemics and pandemics. These models are used to understand the epidemic spreading mechanisms and the transmission dynamics in population. These compartmental models can be divided into two groups i) Population-based models and ii) Agent-based models~\cite{keeling2008modeling}. Population-based models can be subdivided into deterministic, stochastic, and discrete-time models. For this research, we have considered one of the well known compartmental model SEIR. SEIR represents the general interaction among four groups, Susceptible, S (part of the total population who are not yet infected by the disease pathogen), Exposed, E (part of the population who are exposed to the infected individual or been infected by the pathogen but haven't developed any clinical signs), Infected, I (part of the population who are infected by the pathogen and have visible disease symptoms, they can also infect others) and Recovered, R (individuals who survived after being infected but is no longer infectious and has developed a natural immunity to the disease pathogen)~\cite{keeling2008modeling}~\cite{brauer2001math} ~\cite{hethcote2000math}. 
	
\begin{figure}
		\centering
		\includegraphics[height=5.2cm,width=0.9\columnwidth]{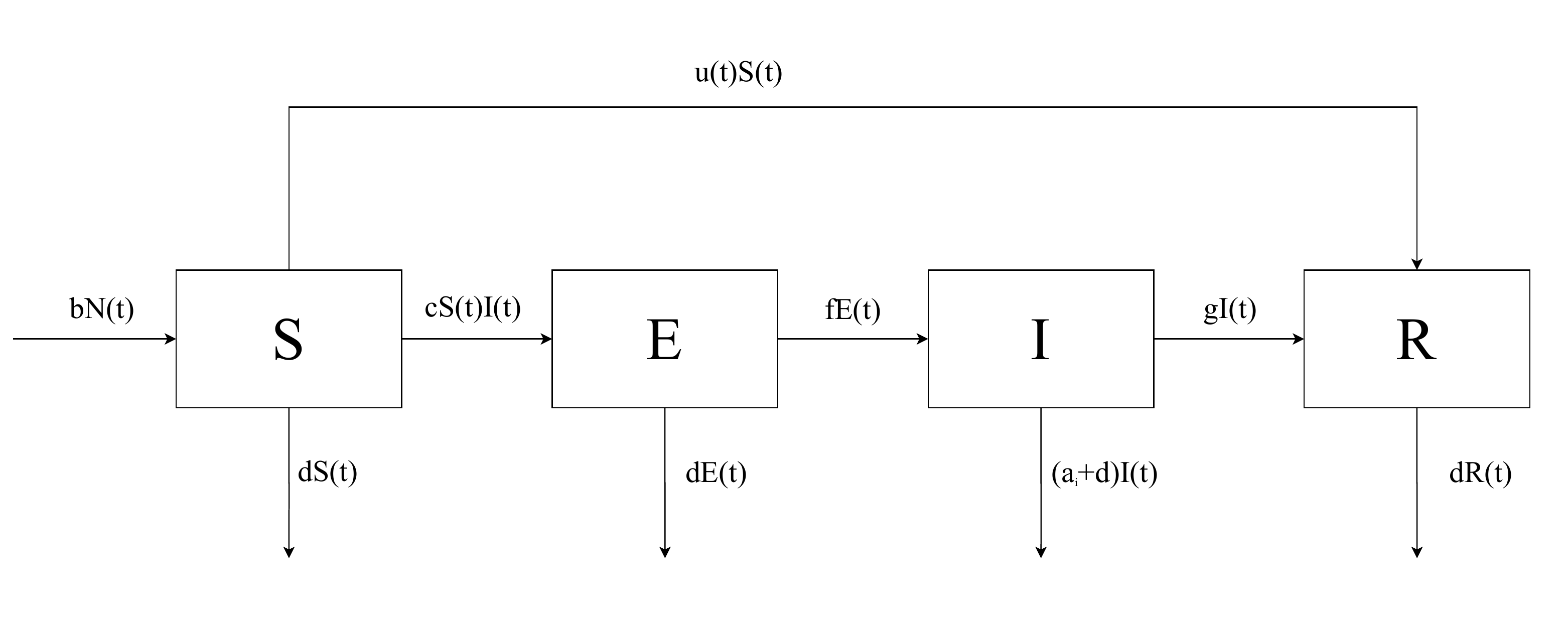}
		\caption{SEIR Compartmental Model}
		\label{fig:SEIRModel}
	\end{figure}

	Let's consider the total population at time \textit{t} is denoted by \textit{N(t)}. Then the total population will divide into four compartments susceptible \textit{S(t)}, Exposed \textit{E(t)}, Infected \textit{I(t)}, and Recovered \textit{R(t)}. Consider we have vaccine available then we will provide it to the susceptible individuals and once they received successful vaccination, they will move to the recovered compartment. Susceptible individuals can be exposed to the disease pathogen and moved to the exposed compartment and eventually move to the infected compartment. People can move out of the infected group by recovering from the disease, or by dying due to infection or other cause. A newly born baby will be added to the susceptible group. Also, people can die due to other causes of each compartment. Consider death rate due to other reasons is \textit{d}, the birth rate is \textit{b}, and death rate due to infection is denoted by \textit{a\textsubscript{i}}. If we denote the rate of vaccination at time \textit{t} by \textit{u(t)} then Fig.~\ref{fig:SEIRModel} shows the classical SEIR model and interaction among the compartments.
	
\begin{figure}[H]
		\centering
		\includegraphics[height=6.9cm,width=0.9\columnwidth]{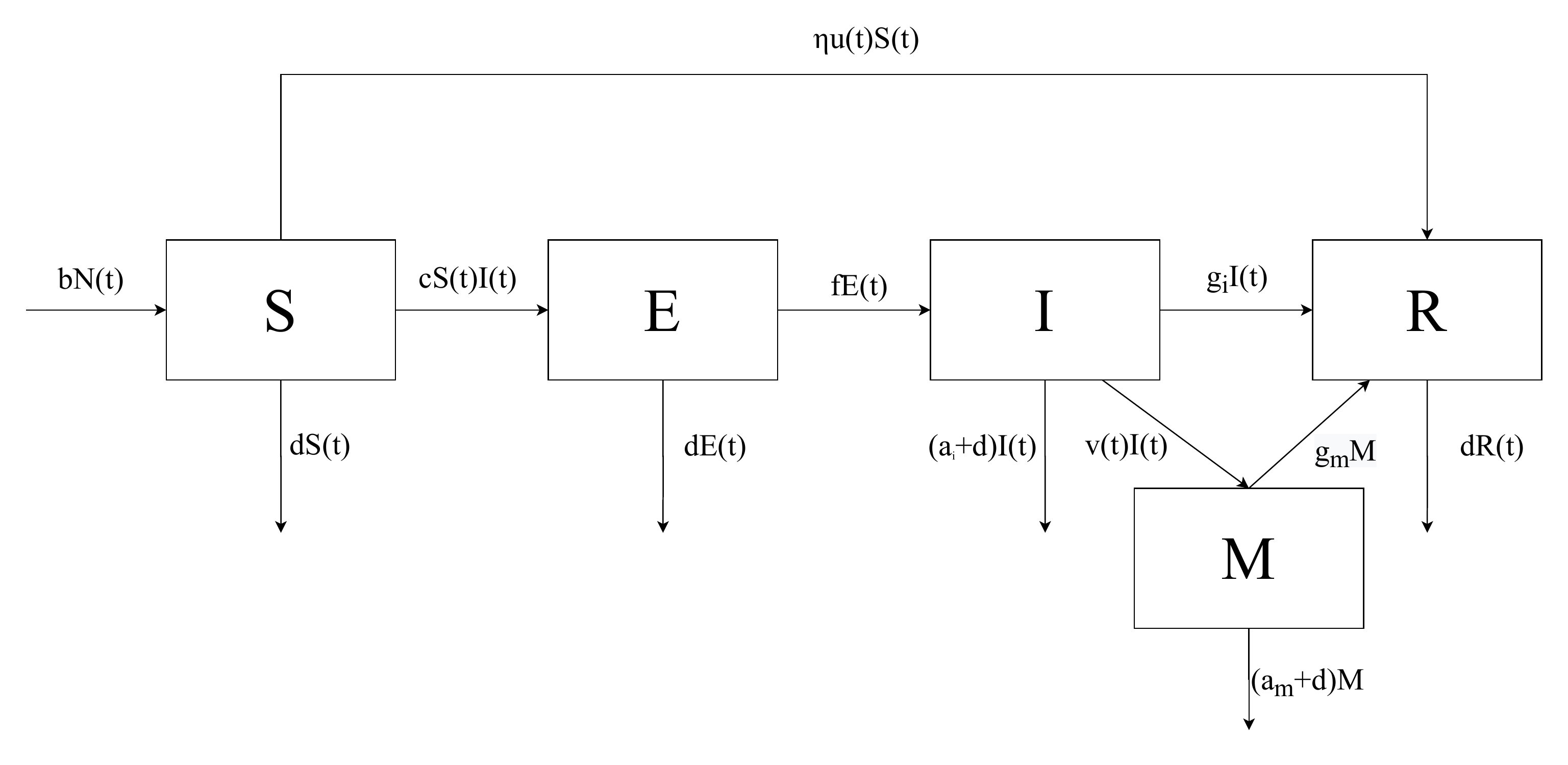}
		\caption{SEIRM Model}
		\label{fig:SEIMRModel}
	\end{figure}

	For our work, we have considered that spreading of the COVID-19 can be stopped by vaccinating the susceptible and offering medical care to those who are infected. We also assume that vaccines have efficiency means not everyone getting the vaccine will not get immune to the virus. The efficiency of the vaccine is denoted by $\eta$. This means if \textit{u(t)S(t)} number of individuals would get vaccine then only \textit{$\eta$u(t)S(t)} people would be immune to the virus and would move to the recovered compartment. Inspired by~\cite{Pinho2016costs}, we have introduced a new compartment in the model denoted by \textit{M(t)}. Individuals who are infected and getting medical care are members of this group. We denote the rate of getting treatment for the infected individuals as \textit{v(t)}. Individuals in this compartment who do not die due to the infection or other causes are also considered immune and moved to the recovered compartment. Considering this new group, the modified model is shown in Fig.~\ref{fig:SEIMRModel} and the system dynamics will be like (\ref{eq:HM1})-(\ref{eq:HM6}). 
	\begin{equation}
	\label{eq:HM1}
	\dot{S}(t) = {bN(t) - dS(t) - cS(t)I(t) - \eta u(t)S(t)},
	\end{equation}
	\begin{equation}
	\label{eq:HM2}
	\dot{E}(t) = {cS(t)I(t) - (f + d)E(t))},
	\end{equation}
	\begin{equation}
	\label{eq:HM3}
	\dot{I}(t) = {fE(t) - (g_i + a_i + d)I(t) - v(t)I(t)},
	\end{equation}
	\begin{equation}
	\label{eq:HM4}
	\dot{R}(t) = {g_iI(t)+g_mM(T) - dR(t) + u(t)S(t)},
	\end{equation}
	\begin{equation}
	\label{eq:HM5}
	\dot{M}(t) = {-(g_m + a_m + d)M(t) + v(t)I(t)},
	\end{equation}
	\begin{equation}
	\label{eq:HM6}
	\dot{N}(t) = {(b - d)N(t) - aI(t)},
	\end{equation}
	Here \textit{S(0)=S\textsubscript{0}}, \textit{E(0)=E\textsubscript{0}}, \textit{I(0)=I\textsubscript{0}}, \textit{R(0)=R\textsubscript{0}}, \textit{M(0)=M\textsubscript{0}}, \textit{N(0)=N\textsubscript{0}}. Where \textit{S\textsubscript{0}}, \textit{E\textsubscript{0}}, \textit{I\textsubscript{0}}, \textit{R\textsubscript{0}}, \textit{M\textsubscript{0}}, \textit{N\textsubscript{0}} are the non-negative initial conditions. Observe that \textit{N(t)} = \textit{S(t)} + \textit{E(t)} + \textit{I(t)} + \textit{M(t)} + \textit{R(t)}. It is important to note that we considered the size of the population constant as the time horizon [0, \textit{T}] is small. Thus the system (\ref{eq:HM1})-(\ref{eq:HM6}) covers almost the same population as the initial population \textit{N\textsubscript{0}} is large and the birth rate could not contribute to the total population in such a short period due to the high death rate during the pandemic. We have normalized the model as in~\cite{Pinho2016costs} considering the percentage of the total population to be 1 at instant \textit{t} which changes the variable to 
	\begin{equation}
	\label{eq:HM7}
	s(t) = \frac{S(t)}{N(t)}, e(t) = \frac{E(t)}{N(t)}, i(t) = \frac{I(t)}{N(t)}, R(t) = \frac{R(t)}{N(t)}, m(t) = \frac{M(t)}{N(t)}
	\end{equation}
	which leads us to the following system dynamics
	\begin{equation}
	\label{eq:HM8}
	\dot{s}(t) = {b-ds(t)-cs(t)i(t)+(a\textsubscript{i}i(t)+a\textsubscript{m}m(t))s(t)-\eta u(t)s(t)},
	\end{equation}
	\begin{equation}
	\label{eq:HM9}
	\dot{e}(t) = {cs(t)i(t)-(f+d)e(t)+(a\textsubscript{i}i(t)+a\textsubscript{m}m(t))e(t)},
	\end{equation}
	\begin{equation}
	\label{eq:HM10}
	\dot{i}(t) = {fe(t)-(g\textsubscript{i}+a\textsubscript{i}+d)i(t)+(a\textsubscript{i}i(t)+a\textsubscript{m}m(t))i(t)-v(t)i(t)},
	\end{equation}
	\begin{equation}
	\label{eq:HM11}
	\dot{r}(t) = {g\textsubscript{i}i(t)+g\textsubscript{m}m(t)-dr(t)+(a\textsubscript{i}i(t)+a\textsubscript{m}m(t))r(t)-\eta u(t)s(t)},
	\end{equation}
	\begin{equation}
	\label{eq:HM12}
	\dot{m}(t) = -(a\textsubscript{m}+g\textsubscript{m}+d)m(t)+{(a\textsubscript{i}i(t)+a\textsubscript{m}m(t))m(t)+v(t)i(t)},
	\end{equation}

	\section{Optimal Control Problem with Mixed Constraints}
	\label{sec:optimal}
	In  an optimal control problem, the control variable is adjusted to achieve a objective or goal. The actual system can have several types of equations namely, ordinary differential equations, partial differential equations, discrete equations, etc. In optimal control definition we denote $u(t)$ as the control and $x(t)$ as the state. The control affects the state function as 
	\begin{equation}
	x'(t) = f(t,x(t),u(t))
	\end{equation}
	where $x'$ denotes the derivative with respect to time t. Both $u(t)$ and $x(t)$ are responsible for goal achievement. Goal is defined by a functional called the objective function. The aim is to find the optimal control
	and corresponding state that achieve the maximum or minimum of our objective function. Objective function can be represented by as following
	\begin{equation}
	Maximize\; or\; Minimize   \quad J(x,u)= \int_{t_{0}}^{t_{f}} L(t,x(t),u(t)) dt
	\end{equation}
	where $L$ is running cost , $ t_i$ is the initial time when the control starts and $t_f$ is the final time which is either fixed or free depending on the goal.
	
	According to the model described section~\ref{sec:math} we have considered the following optimal control problem with control state constraints
	\begin{equation}
	A_{1}\begin{cases}
	\label{eq:A1}
	Minimize   \quad J\textsubscript{1}(x,u,v)= \int_{0}^{T} (Qi(t)+Wu(t)+Yv(t)) dt\\
	Subject\; to\\
	\dot{s}(t) = {b-ds(t)-cs(t)i(t)+(a\textsubscript{i}i(t)+a\textsubscript{m}m(t))s(t)-\eta u(t)s(t)}\\
	\dot{e}(t) = {cs(t)i(t)-(f+d)e(t)+(a\textsubscript{i}i(t)+a\textsubscript{m}m(t))e(t)}\\
	\dot{i}(t) = {fe(t)-(g\textsubscript{i}+a\textsubscript{i}+d)i(t)+(a\textsubscript{i}i(t)+a\textsubscript{m}m(t))i(t)-v(t)i(t)}\\
	\dot{m}(t) = -(a\textsubscript{m}+g\textsubscript{m}+d)m(t)+{(a\textsubscript{i}i(t)+a\textsubscript{m}m(t))m(t)+v(t)i(t)}\\
	u(t) \;\,\epsilon \quad[0,1] \qquad for\;\, a.e.\quad t \;\,\epsilon \;\;[0,T]\\
	v(t) \;\,\epsilon \quad[0,1] \qquad for\;\, a.e.\quad t \;\,\epsilon \;\;[0,T]
	\end{cases}
	\end{equation}	
	Here $Q$,$W$,$Y$ are constant and called infected, vaccination and treatment weight parameters respectively. They are used to focus the optimization from the aspect of infection, vaccination or medication. We have eliminated the equation for evaluation of the percentage recovered $r(t)$ as $s(t) + e(t) + i(t) + r(t) = 1$ for all time $t$. From problem $A_1$ it is clear that this controller will minimize the number of infected individual while using distribution the optimal number of vaccine possible. What if we have a limited vaccine production or limited logistic support, which confines us to use an almost fixed number of vaccine available. To represent that we have introduced a mixed control-state constraint of the form
	\begin{equation}
	u(t)s(t) \leq V
	\end{equation} 
	where $V$ is the upper bound of the vaccine available at each time instant $t$. Then the total optimal vaccine scheduling problem becomes as follow
	\begin{equation}
	A_{2}\begin{cases}
	\label{eq:A2}
	Minimize   \quad J_{2}(x,u,v)= \int_{0}^{T} (Qi(t)+Wu(t)+Yv(t)) dt\\
	Subject\; to\\
	\dot{s}(t) = {b-ds(t)-cs(t)i(t)+(a\textsubscript{i}i(t)+a\textsubscript{m}m(t))s(t)-\eta u(t)s(t)}\\
	\dot{e}(t) = {cs(t)i(t)-(f+d)e(t)+(a\textsubscript{i}i(t)+a\textsubscript{m}m(t))e(t)}\\
	\dot{i}(t) = {fe(t)-(g\textsubscript{i}+a\textsubscript{i}+d)i(t)+(a\textsubscript{i}i(t)+a\textsubscript{m}m(t))i(t)-v(t)i(t)}\\
	\dot{m}(t) = -(a\textsubscript{m}+g\textsubscript{m}+d)m(t)+{(a\textsubscript{i}i(t)+a\textsubscript{m}m(t))m(t)+v(t)i(t)}\\
	u(t)s(t) \leq V\\
	u(t) \;\,\epsilon \quad[0,1] \qquad for\;\, a.e.\quad t \;\,\epsilon \;\;[0,T]\\
	v(t) \;\,\epsilon \quad[0,1] \qquad for\;\, a.e.\quad t \;\,\epsilon \;\;[0,T]
	\end{cases}
	\end{equation}
	
	\section{RBF-Galerkin Method for Optimal Control Problem}
	\label{sec:RBF}
	In 1968 Hardy~\cite{Hardy1971Multiquadric} developed Radial Basis Function (RBF) for function approximation. RBF is a real-valued function whose value depends only on the distance from a fixed point called center point.
	\begin{equation}
	\label{eq:RBF}
	\phi(x,c) = \phi(\lVert {x-c} \rVert)
	\end{equation}
	where $\phi$ is the RBF function and c is the center. Any function that satisfies (\ref{eq:RBF}) is called the RBF function. As RBFs are multivariate functions, it can be applied in any direction~\cite{buhmann2003radial}. RBF-Galerkin method is based on interpolation of global RBFs on an arbitrary set of collocation points. A various set of collocation points can be chosen arbitrarily for discretization. For example, a set of ChebyshevGauss (CG), Chebyshev-Gauss-Lobatto (CGL), Legendre-Gauss (LG), and Legendre-Gauss-Lobatto (LGL) nodes, each could be selected as a set of unequally spaced orthogonal nodes to discretize the problem~\cite{Williams2009Hermite}. Similarly, can be used a set of equally-spaced nodes in the time for discretization. Among the common sets of collocation points, two most popular choices are sets of CGL and LGL nodes distributed over the interval [-1,1]. The first set minimizes the max-norm of the interpolation error, while the second one minimizes the $L^2$-norm of the interpolation error. 
	
	According to~\cite{Mirinejad2015Individualized}, an approach based on the RBF-Galerkin method is proposed to solve the optimal control problem $A_1$ and $A_2$. To solve the two optimization problems $s(t)$, $e(t)$, $i(t)$, $m(t)$, $u(t)$, and $v(t)$ are approximated using $k$ global RBF functions within time horizon [$0$, $T$] as following
	
	\begin{align}
	s(t) &= s^{R}(t) = \sum_{j=1}^{k}\alpha _{j} \phi(\lVert {t-t_{j}} \rVert) = \sum_{j=1}^{k} \alpha _{j} \phi _{j}(t)\\
	e(t) &= e^{R}(t) = \sum_{j=1}^{k}\beta _{j} \phi(\lVert {t-t_{j}} \rVert) = \sum_{j=1}^{k} \beta _{j} \phi _{j}(t)\\
	i(t) &= i^{R}(t) = \sum_{j=1}^{k}\alpha _{j} \phi(\lVert {t-t_{j}} \rVert) = \sum_{j=1}^{k} \alpha _{j} \phi _{j}(t)\\
	m(t) &= m^{R}(t) = \sum_{j=1}^{k}\lambda _{j} \phi(\lVert {t-t_{j}} \rVert) = \sum_{j=1}^{k} \lambda _{j} \phi _{j}(t)\\
	u(t) &= u^{R}(t) = \sum_{j=1}^{k}\mu _{j} \phi(\lVert {t-t_{j}} \rVert) = \sum_{j=1}^{k} \mu _{j} \phi _{j}(t)\\
	v(t) &= v^{R}(t) = \sum_{j=1}^{k}\sigma _{j} \phi(\lVert {t-t_{j}} \rVert) = \sum_{j=1}^{k} \sigma _{j} \phi _{j}(t)
	\end{align}
	
	where $s^R(t)$, $e^R(t)$,$i^R(t)$,$m^R(t)$,$u^R(t)$ and $v^R(t)$ denotes the RBF approximation of $s(t)$, $e(t)$,$i(t)$,$m(t)$,$u(t)$ and $v(t)$ respectively. $\phi_{j}$ is the RBF basis function and $\alpha_{j}$, $\beta_{j}$, $\gamma_{j}$, $\lambda_{j}$, $\mu_{j}$, $\sigma_{j}$ are RBF weights.The problem is then discretized using Legendre-Gauss-Lobatto (LGL) nodes. LGL nodes $\tau_{j}$ , j =1, 2,...,k are orthogonal nodes distributed
	over the interval [-1,1] given by $\tau_{1}= -1$ and $\tau_{k} = 1$ for $2\leq j \leq K-1$ are the zeros of $\dot{P}_{K-1}$, the derivative of the Legendre polynomial of degree $K-1$, $P_{K-1}$. The integral cost functions of $A_{1}$ and $A_{2}$ are also approximated by the Gauss-Lobatto quadrature as
	
	\begin{equation}
	J_{1} = \frac{T}{2} \sum_{j-1}^{k}(w_{j}(Qi(t)+Wu(t)+Tv(t)))
	\end{equation}
	\begin{equation}
	J_{2} = \frac{T}{2} \sum_{j-1}^{k}(w_{j}(Qs(t)+Wu(t)+Tv(t)))
	\end{equation}
	
	where $w_{j}$ are LGL weights corresponding to LGL nodes, $\tau_{j} \, \epsilon \,[-1,1]$, given by
	\begin{equation}
	w_{j} = \frac{2}{K(K-1)[P_{K-1}(\tau_{j})]^2}  \qquad\text{where $j=1,2,....K$}
	\end{equation}
	Then the differential equation in (\ref{eq:A1}) and (\ref{eq:A2}) are converted into algebraic equations using the Gaussian differentiation matrix~\cite{Mirinejad2015radial} to translate the original Problem $A_{1}$ and $A_{2}$ into Non-Linear Programming (NLP) optimization problem. The NLP problem has the general form 
	\begin{align}
	\text{minimize} \qquad &f(y)\\\nonumber
	\text{Subject to} \qquad&lb\leq g(y)\leq ub
	\end{align}
	
	Here \textit{y} is the vector of decision variables, \textit{g(y)} is a set of equality and inequality constraints, and lb and ub denote lower-bound and upper-bound vectors. The NLP problem can then be straightforwardly solved using NLP solvers. For this work, SNOPT ~\cite{Gill2002SNOPT}, a sparse NLP solver, is used to find the optimum solution.
	
	\section{Numerical Results}
	\label{sec:numerical solution}
	For the numerical solution, two different initial value datasets have been used. One dataset coincided with those in ~\cite{Neilan2010Into} to solve problem $A_1$ and $A_2$ initially. Then we have used the data from Worldometer~\cite{worldometer} for New York State as the initial values and constants to solve for vaccination policy. Results are presented in two subsections, ~\ref{initial} discusses the results found for the hypothetical dataset, and ~\ref{newyork} we have discusses the results for the New York State, USA dataset.
	
	\subsection{Results for Benchmark (Synthetic) Data}
	\label{initial}
\begin{figure}[H]
		\centering
		\includegraphics[width=1\columnwidth]{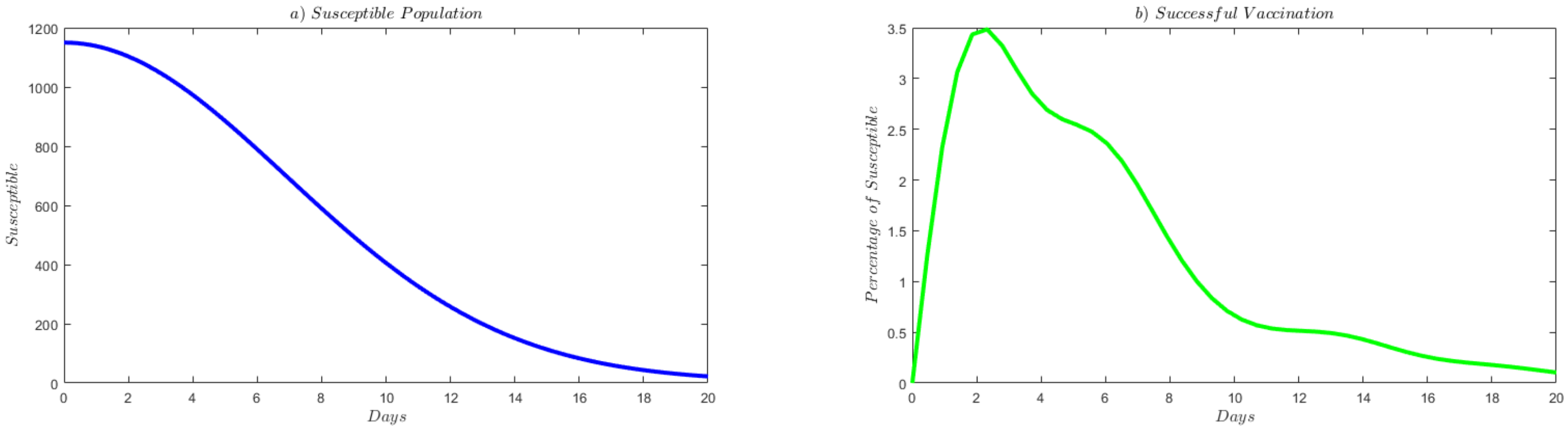}
		\caption{Susceptible and Vaccination Trajectory for problem $A_{1}$}
		\label{fig:data1}
	\end{figure}

\begin{figure}[H]
		\centering
		\includegraphics[width=0.98\columnwidth]{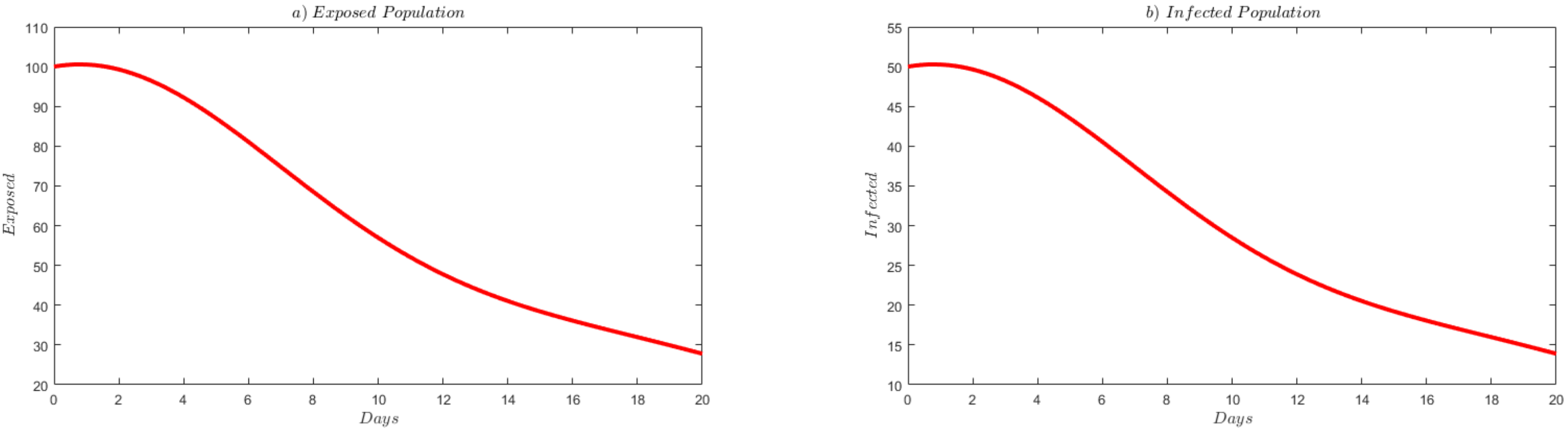}
		\caption{Trajectory of Exposed and Infected Population for problem $A_{1}$}
		\label{fig:data2}
	\end{figure}

	We have utilized the data presented in~\cite{Neilan2010Into} to test our models and solver that uses RBF-Galerking method.This allowed to compare our results with other researchers and to find suitable parameters for the RBF functions specific for the problem $A_1$ and $A_2$. The initial data are shown in Table \ref{tab:table2}.
	\begin{table}[H]
		\begin{center}
			\caption{Parameters with their clinically approved values and constants as in~\cite{Neilan2010Into}}
			\label{tab:table2}
			\begin{tabular}{lll} 
				\hline
				\textbf{Parameter} & \textbf{Description} & \textbf{Value} \\
				\hline 
				\vspace{2pt}
				$b$ & natural birth rate  &  $0.525$\\
				\vspace{2pt}
				$d$  & natural death rate &  $0.5$\\
				\vspace{2pt}
				$c$ & incidence coefficient &  $0.001$\\
				\vspace{2pt}
				$f$ & exposed to infectious rate &  $0.5$\\
				\vspace{2pt}
				$g_{i}$  & recovery rate of those infected &  $0.1$\\
				\vspace{2pt}
				$g_{m}$  & recovery rate of those treated &  $0.1$\\
				\vspace{2pt}
				$a_{i}$  & infection induced death rate &  $0.2$\\
				\vspace{2pt}
				$a_{m}$  & treatment induced death rate &  $0.2$\\
				\vspace{2pt}
				$\eta$  & efficiency of vaccine &  $0.8$\\
				\vspace{2pt}
				$S_{0}$ & initial susceptible population & $1000$\\
				\vspace{2pt}
				$E_{0}$ & initial exposed population & $100$\\
				\vspace{2pt}
				$I_{0}$ & initial infected population & $50$\\
				\vspace{2pt}
				$R_{0}$ & initial recovered population & $15$\\
				\vspace{2pt}
				$N_{0}$ & initial population & $1165$\\
				\vspace{2pt}
				$Q$ & infected weight parameter & $100$\\
				\vspace{2pt}
				$W$ & vaccination weight parameter & $30$\\
				\vspace{2pt}
				$Y$ & treatment weight parameter & $1$\\
				\vspace{2pt}
				$T$ & number of days & $20$\\
				\hline
			\end{tabular}
		\end{center}
	\end{table} 
	We have used the RBF-Galerkin method to turn the optimal control problem into an NLP problem and used the SNOPT~\cite{Gill2002SNOPT} tool in Matlab to solve the NLP problem. We have used 100 nodes grid with step size $\Delta t = T /100 = 0.2$. We have considered the upper bound of available vaccine $V=125$ which is 10\% of the Susceptible and Exposed Population. Fig.~\ref{fig:data1} and Fig.~\ref{fig:data2} show the results for the problem $A_{1}$. In Fig.~\ref{fig:data1}b the optimal vaccination policy is presented which lower the susceptible population to an acceptable low value (see Fig.~\ref{fig:data1}a) as the susceptible population cannot go to zero. The lowest susceptible population can go is the birthrate of the total population as the newborn children are considered not immune to the virus. But here the susceptible population is still higher than the birthrate. Meanwhile, the exposed and infected population also go down (see Fig.~\ref{fig:data2}a and Fig.~\ref{fig:data2}b). However, as the Susceptible Population does not reduce to the birth rate and the incidence rate $c$ is high, the Exposed and Infected population haven't reach zero in the defined time horizon. In this problem we haven't put any bounds on the number of available vaccine. Due to the use of higher weight parameter for the vaccination, the solver optimizes the number of vaccine provided each day, thus the vaccine policy shown in the Fig.~\ref{fig:data1}b starts at zero for time $t_{0}$ and graphs show the optimal possible vaccine policy.
	
\begin{figure}[H]
		\centering
		\includegraphics[width=1\columnwidth]{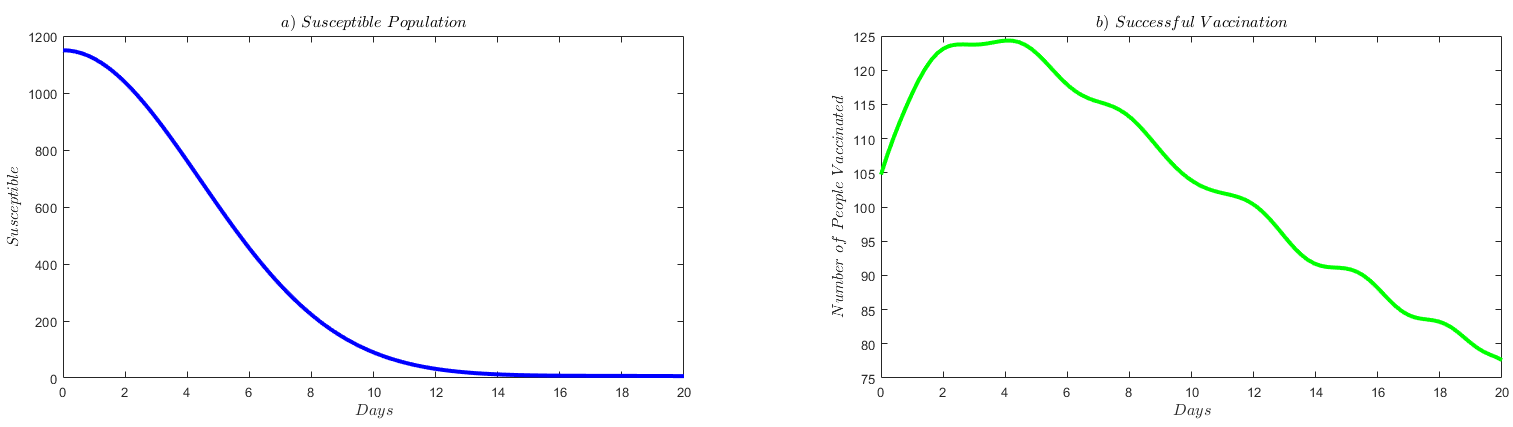}
		\caption{Susceptible and Vaccination Trajectory for problem $A_{2}$}
		\label{fig:data3}
	\end{figure}

\begin{figure}[H]
		\centering
		\includegraphics[width=1\columnwidth]{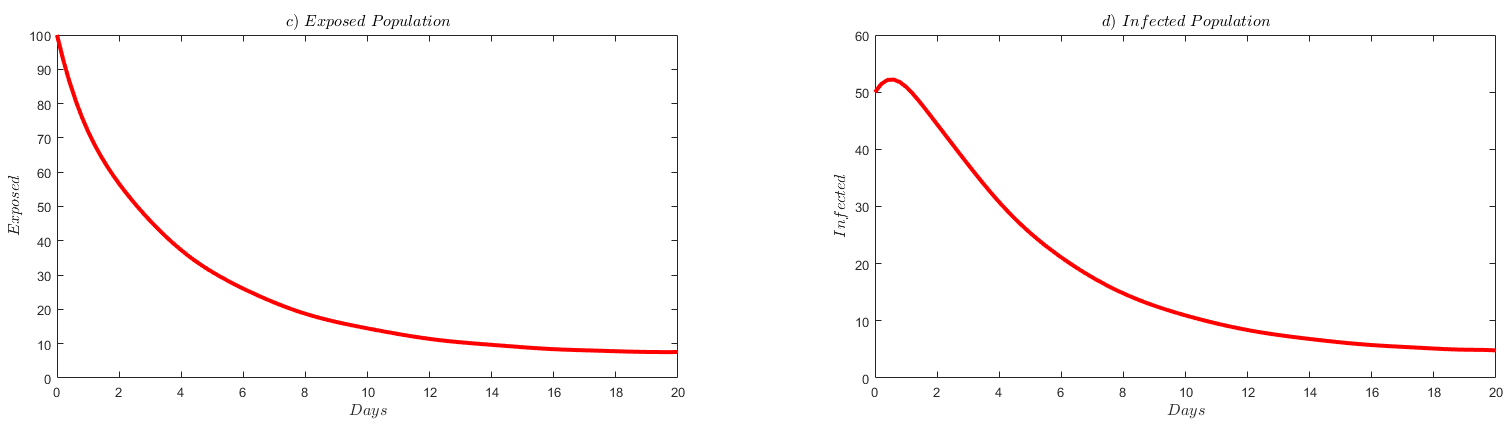}
		\caption{Trajectory of Exposed and Infected Population for problem $A_{2}$}
		\label{fig:data4}
	\end{figure}

	Fig.~\ref{fig:data3} and Fig.~\ref{fig:data4} shows the results for problem $A_{2}$. As we have set a limit of $125$ vaccines available per day, we have shown the vaccinated population as numbers rather than percentages. Fig.~\ref{fig:data3}b shows the vaccine distribution. It is noticeable that the values are always under $125$. As this is the optimal policy with an upper bound defined so the number doesn't start from zero but from a large value. Here we are not considering the leftover vaccines from the following day in the problem. The Susceptible Population curve of Fig.~\ref{fig:data3}a is now steeper than before and it reaches the birth rate in $17$ days. As the susceptible population is dropping faster the exposed and infected number also goes down to near zero in the defined time horizon.

	\subsection{Results for New York, USA Data}
	\label{newyork}
	For real world data we have taken the data for New York State, USA considered as the initial epicenter of COVID-19. Finding the optimal vaccine schedule in small time span for New York will be a challenging task. The initial data was taken on May 12, 2020, is shown in Table \ref{tab:table3}. As the values are large, they are taken in percentage to make calculations faster and easier.
	
\begin{figure}[H]
		\centering
		\includegraphics[width=0.98\columnwidth]{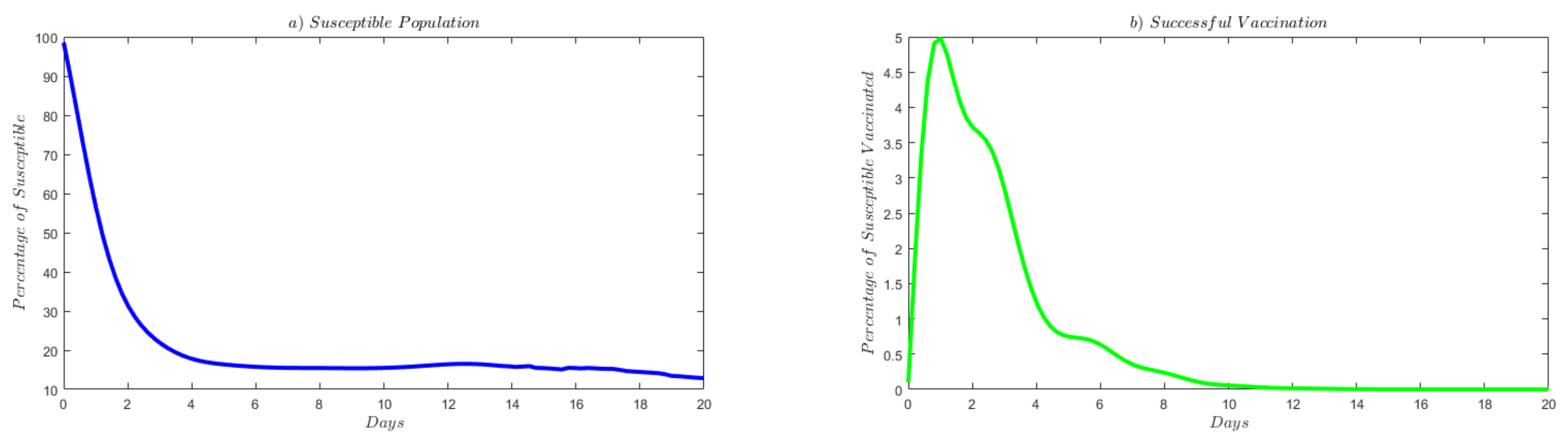}
		\caption{Susceptible and Vaccination Trajectory for New York using problem $A_{1}$}
		\label{fig:data5}
	\end{figure}

\begin{figure}[H]
		\centering
		\includegraphics[width=0.98\columnwidth]{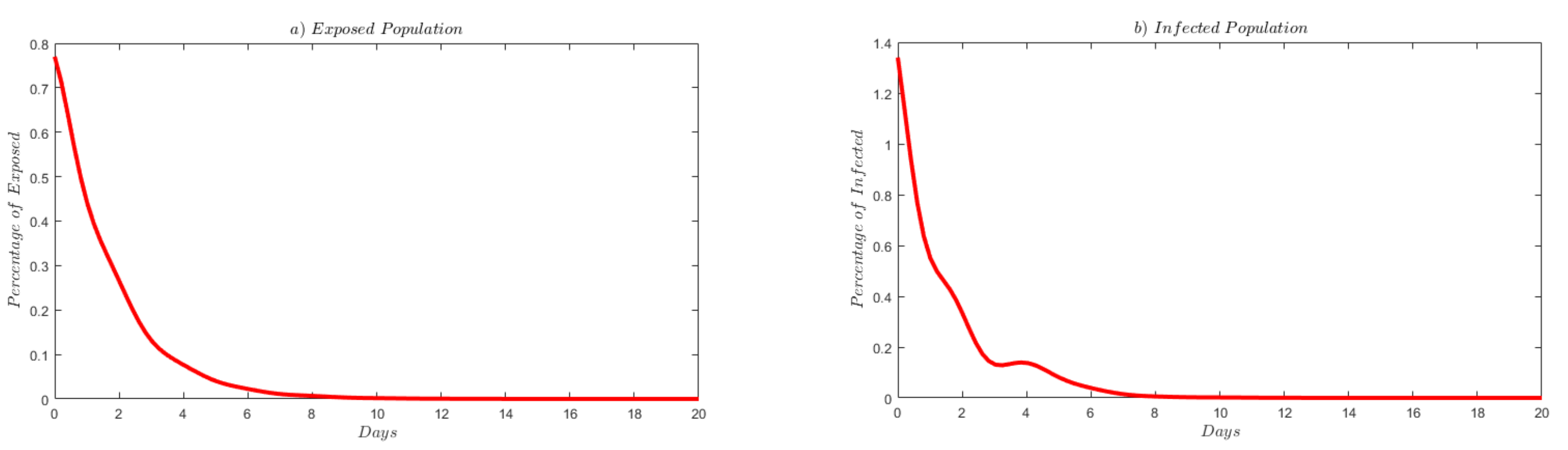}
		\caption{Trajectory of Exposed and Infected Population of New York using problem $A_{1}$}
		\label{fig:data6}
	\end{figure}

\begin{figure}[H]
		\centering
		\includegraphics[width=0.97\columnwidth]{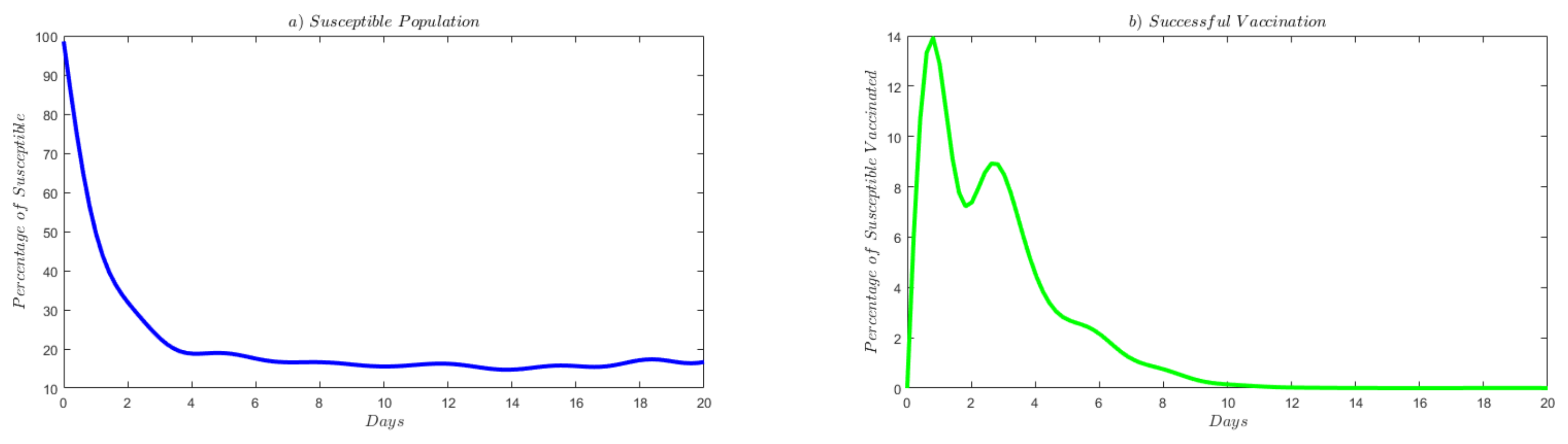}
		\caption{Susceptible and Vaccination Trajectory for New York using problem $A_{2}$}
		\label{fig:data7}
	\end{figure}

\begin{figure}[H]
		
		\centering
		\includegraphics[width=0.97\columnwidth]{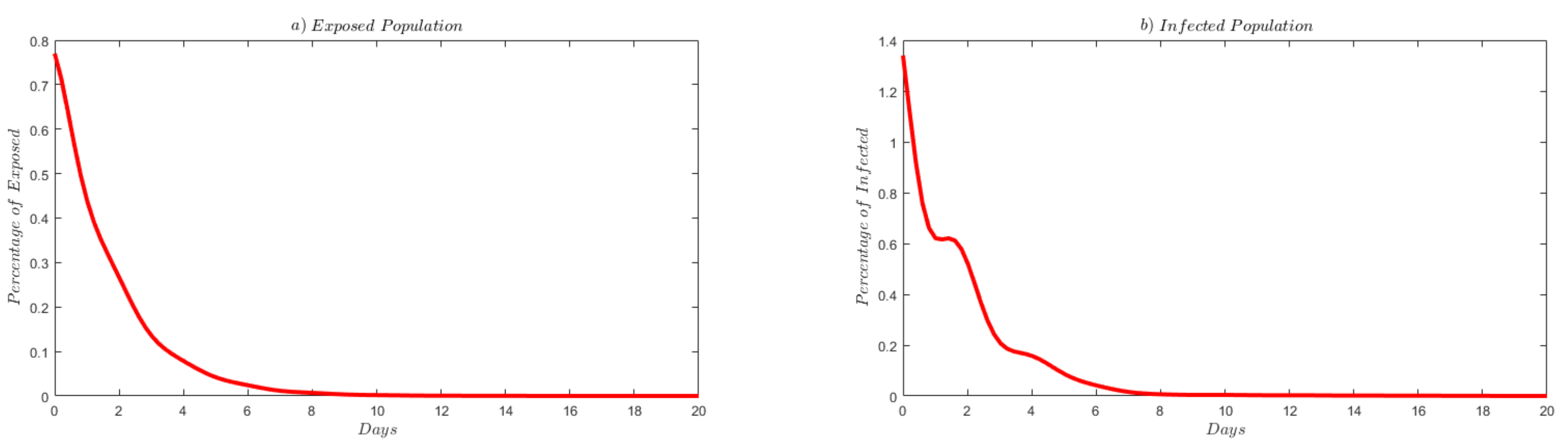}
		\caption{Trajectory of Exposed and Infected Population of New York using problem $A_{2}$}
		\label{fig:data8}
	\end{figure}

	\begin{table}[H]	
		\begin{center}
			\caption{Parameter Values for New York State, USA}
			\label{tab:table3}
			\begin{tabular}{lll} 
				\hline
				\textbf{Parameter} & \textbf{Description} & \textbf{Value} \\
				\hline 
				\vspace{2pt}
				$b$ & natural birth rate  &  $11.5$\\
				\vspace{2pt}
				$d$  & natural death rate &  $7.8$\\
				\vspace{2pt}
				$c$ & incidence coefficient &  $0.11$\\
				\vspace{2pt}
				$f$ & exposed to infectious rate &  $0.5$\\
				\vspace{2pt}
				$g_{i}$  & recovery rate of those infected &  $16.84\%$\\
				\vspace{2pt}
				$g_{m}$  & recovery rate of those treated &  $75.45\%$\\
				\vspace{2pt}
				$a_{i}$  & infection induced death rate &  $3\%$\\
				\vspace{2pt}
				$a_{m}$  & treatment induced death rate &  $0.5\%$\\
				\vspace{2pt}
				$\eta$  & Efficiency of vaccine &  $0.8$\\
				\vspace{2pt}
				$S_{0}$ & initial susceptible population & $97.57\%$\\
				\vspace{2pt}
				$E_{0}$ & initial exposed population & $0.77\%$\\
				\vspace{2pt}
				$I_{0}$ & initial infected population & $1.34\%$\\
				\vspace{2pt}
				$R_{0}$ & initial recovered population & $0.32\%$\\
				\vspace{2pt}
				$Q$ & infected weight parameter & $100$\\
				\vspace{2pt}
				$W$ & vaccination weight parameter & $30$\\
				\vspace{2pt}
				$Y$ & treatment weight parameter & $1$\\
				\vspace{2pt}
				$T$ & number of days & $20$\\
				\hline
			\end{tabular}
		\end{center}
	\end{table} 
	The Fig.~\ref{fig:data5} and Fig.~\ref{fig:data6} show the result for New York when problem $A_1$ is considered. The Susceptible Population curve is much steeper and reaches near the birth rate value faster. The vaccine distribution curve has peak in the beginning and reaches a saturation point in $9$ days. Due to the early drop in the susceptible curve and lower incident coefficient the percentage of exposed and infected goes down to almost zero in 8 days (as shown in Fig.~\ref{fig:data6}). As the infected weight parameter $Q$ is high, when the percentage of Infected individuals decreases the vaccine distribution is also reduced (as shown in Fig.~\ref{fig:data5}).  
	\newpage
	For problem $A_{2}$ we have used a limit of $15\%$ of Susceptible Population to bound the maximum vaccine distribution. This gives us the results shown in Fig.~\ref{fig:data7}~\ref{fig:data8}. Limiting the upper bound gives a higher initial distribution of vaccination as in Fig.~\ref{fig:data7}b. Saturation takes a similar time as in the optimal control formulation $A_{1}$. The percentage of Susceptible Population goes down to the birth rate in $6$ days. There is a sudden increase in the vaccine distribution at day $3$. This is due to an increase in the infected population around day $2$ (see Fig.~\ref{fig:data8}b). At day $1$ there is a drop in infected individuals which leads to a drop in vaccination at day $2$, that makes slop of the susceptible curve flatter than the day $1$. Which means less people are getting vaccine and moving out of the susceptible, in other words more people are getting exposed and eventually getting infected. That makes the infected curve to have a flat region on day $2$. 
	
	Both the optimal control formulation of optimal vaccine scheduling $A_1$ and $A_2$ aim to reduce the Infected Population while utilizing the optimal vaccination possible. However, this objective function can could be set up differently. For instance, we can emphasis to reduce the susceptible rather than the infected population, or we can give more emphasis on medication. 
	\section{Conclusion}
	\label{sec:conclusion}
	Vaccine distribution to the large population is always a challenge. As numerous companies are developing vaccines for the COVID-19, dynamic vaccine distribution policy will be of high importance when the vaccines will become available. In this study, we have considered widely accepted infectious disease model and solved it for limited vaccine availability. We have formulated the scheduling of vaccination as two optimal control problems and then presented numerical solutions with two different datasets, one with real-world COVID-19 related data for New York.USA. We have used an RBF-Galerkin method for solving the two optimal control formulations. The formulations presented to reduce the infected population while distributing the vaccine at optimal rate. As the distribution largely depends on the infected population it can sometimes ignore the population of other groups. However, the optimal control problems are dynamic and they can be formulated to met any other requirements needed for efficient distribution.

	\bibliographystyle{apalike}
	\bibliography{reference}

\begin{thebibliography}{10}

\bibitem{huang2020clinical}
Chaolin Huang, Yeming Wang, Xingwang Li, Lili Ren, and Jianping Zhao.
\newblock Clinical features of patients infected with 2019 novel coronavirus in
  wuhan, china.
\newblock {\em The Lancet}, 395(10223):497–506, 2020.

\bibitem{lu2020out}
Hongzhou Lu, Charles~W Stratton, and Yi-Wei Tang.
\newblock Outbreak of pneumonia of unknown etiology in wuhan china: the mystery
  and the miracle.
\newblock {\em Journal of Medical Virology}, 92(4):401–402, 2020.

\bibitem{ictv}
International committee on taxonomy of viruses (ictv) website.
\newblock \url{https://talk.ictvonline.org/}.

\bibitem{who}
World health organization (who) website.
\newblock
  \url{https://www.who.int/emergencies/diseases/novel-coronavirus-2019/technical-guidance/}.

\bibitem{li2020early}
Qui Li, Xuhua Guan, Peng Wu, and Xiaoye Wang.
\newblock Early transmission dynamics in wuhan, china, of novel
  coronavirus-infected pneumonia.
\newblock {\em New England Journal of Medicine}, 382(13):1199--1207, 2020.

\bibitem{lauer2020incu}
Stephen~A Lauer, Kyra~H Grantz, Qifang Bi, and Forrest~K Jones.
\newblock The incubation period of coronavirus disease 2019 (covid-19) from
  publicly reported confirmed cases: Estimation and application.
\newblock {\em Annals of Internal Medicine}, 172(9):577--582, 2020.

\bibitem{worldometer}
Worldometer website.
\newblock \url{https://www.worldometers.info/coronavirus/}.

\bibitem{who2}
World health organization (who) website.
\newblock
  \url{https://www.who.int/news-room/detail/30-01-2020-statement-on-the-second-meeting-of-the-international-health-regulations-(2005)-emergency-committee-regarding-the-outbreak-of-novel-coronavirus-(2019-ncov)/}.

\bibitem{CNNVaccine}
9 pharmaceutical companies racing for a covid-19 vaccine.
\newblock
  \url{https://www.forbes.com/sites/moneyshow/2020/06/16/9-pharmaceutical-companies-racing-for-a-covid-19-vaccine/#11aecba976ad}.

\bibitem{kim2016constrained}
Jungeun Kim, Hee-Dae Kwon, and Jeehyun Lee.
\newblock Constrained optimal control applied to vaccination for influenza.
\newblock {\em Computers and Mathematics with Applications}, 71(11):2313--2329,
  2016.

\bibitem{Pinho2014Optimal}
Maria Do~Rosario De~Pinho, Igor Kornienko, and Helmut Maurer.
\newblock Optimal control of a seir model with mixed constraints and $l^{1}$
  cost.
\newblock In {\em 11th Portuguese Conference on Automatic Control}. Springer,
  2014.

\bibitem{Pinho2016costs}
Maria Do~Rosario De~Pinho and Filipa~Nunes Nogueira.
\newblock Costs analysis for the application of optimal control to seir
  normalized models.
\newblock {\em International Federation of Automatic Control (IFAC)},
  21(27):122--127, 2018.

\bibitem{Pinho2016application}
Maria Do~Rosario De~Pinho and Filipa~Nunes Nogueira.
\newblock On application of optimal control to seir normalized models: Pros and
  cons.
\newblock {\em Mathematical Biosciences and Engineering}, 14(1):111--126, 2016.

\bibitem{libotte2020determination}
Gustavo~Barbosa Libotte, Fran~Sérgio Lobato, Gustavo~Mendes Platt, and
  Antônio~José da~Silva~Neto.
\newblock Determination of an optimal control strategy for vaccine
  administration in covid-19 pandemic treatment, 2020.

\bibitem{Mirinejad2016RBF}
Hossein Mirinejad and Tamer Inanc.
\newblock An rbf collocation method for solving optimal control problems.
\newblock {\em Robotics and Autonomous Systems}, 87, 2016.

\bibitem{keeling2008modeling}
Matt~J Keeling and Pejman Rohani.
\newblock {\em Modeling Infectious Diseases in Humans and Animals}.
\newblock Princeton University Press, 2007.

\bibitem{brauer2001math}
Fred Brauer and Carlos~Castillo Chávez.
\newblock {\em Mathematical Models in Population Biology and Epidemiology}.
\newblock Springer, 2001.

\bibitem{hethcote2000math}
Herbert~W. Hethcote.
\newblock The mathematics of infectious diseases.
\newblock {\em Society for Industrial and Applied Mathematics},
  42(4):599–653, 2000.

\bibitem{Hardy1971Multiquadric}
Rolland~L. Hardy.
\newblock Multiquadric equations of topography and other irregular surfaces.
\newblock {\em Journal of Geophysical Research}, 76(8):1905–1915, 1971.

\bibitem{buhmann2003radial}
Martin~D. Buhmann.
\newblock {\em Radial Basis Functions: Theory and Implementations}.
\newblock Cambridge University Press, 2003.

\bibitem{Williams2009Hermite}
Paul Williams.
\newblock Hermite–legendre–gauss–lobatto direct transcription in
  trajectory optimization.
\newblock {\em Journal of Guidance Control and Dynamics - J GUID CONTROL
  DYNAM}, 32(4):1392--1395, 2009.

\bibitem{Mirinejad2015Individualized}
Hossein Mirinejad and Tamer Inanc.
\newblock Individualized anemia management using a radial basis function
  method.
\newblock In {\em IEEE Great Lakes Biomedical Conference, GLBC - Proceedings}.
  IEEE, 2015.

\bibitem{Mirinejad2015radial}
Hossein Mirinejad and Tamer Inanc.
\newblock A radial basis function method for direct trajectory optimization.
\newblock In {\em American Control Conference (ACC)}. IEEE, 2015.

\bibitem{Gill2002SNOPT}
Philip Gill, Walter Murray, and Michael Saunders.
\newblock Snopt: An sqp algorithm for large-scale constrained optimization.
\newblock {\em SIAM Journal on Optimization}, 12(4):979--1006, 2002.

\bibitem{Neilan2010Into}
Rachael~Miller Neilan and Suzanne Lenhart.
\newblock An introduction to optimal control with an application in disease
  modeling.
\newblock {\em DIMACS Series in Discrete Mathematics}, 75:67–81, 2010.

\end{thebibliography}
	
	%TC:endignore
	
	% Word count
	
\end{document}